# Evolutionary adaptation through bet-hedging in finite populations under fluctuating environments

Marco Mendívil-Carboni,[1] Juan J. Mazo,[1] and Luis Dinis[1]

[1] *Grupo Interdisciplinar de Sistemas Complejos (GISC) and Departamento de Estructura de la Materia, Física Térmica y Electrónica, Universidad Complutense de Madrid, 28040 Madrid, Spain*

Natural populations evolve under fluctuating environments and limited resources, yet it is unclear how these factors jointly shape adaptation. Here, we introduce a minimal stochastic model of a population undergoing a birth-death process in a randomly switching environment, where individuals inherit phenotypic strategies that determine the distribution of phenotypes across generations. First, we confirm that environmental variability can favor phenotypic diversity (bet-hedging), and that the strategy maximizing the average growth rate generally differs from the one minimizing extinction risk. We then demonstrate that evolution drives populations toward strategies that balance these competing objectives. Furthermore, we show that population size modulates this evolutionary outcome: large populations converge to the growth-maximizing strategy, whereas smaller ones evolve toward more resilient strategies. Finally, we provide a simple heuristic approximation that captures these results, clarifying the interplay between growth, extinction, and mutation.

## I. INTRODUCTION AND OBJECTIVES

Adaptation in fluctuating environments is a central problem in evolutionary biology [1–13]. Natural populations rarely experience constant conditions: instead, they face environments that change *unpredictably* over time, often on evolutionary timescales. In such contexts, if sensing the changes in the environment is infeasible or too costly, it is well established in the literature that bet-hedging (i.e. investing only a fraction of the population in each possible phenotype) offers a good alternative to adapt to this fluctuating environment [3, 5, 10, 12–14].

In these circumstances, our current understanding of evolutionary adaptation suggests that populations tend to evolve toward the strategy which maximizes their mean growth rate

$$\langle \mu \rangle = \left\langle \frac{1}{N(t)} \frac{dN}{dt}(t) \right\rangle \tag{1}$$

as first argued in [1]. More precisely, [1] argues that what is maximized is the geometric mean of the offspring number $\mu'$ (where $N(t+\Delta t) = \mu' \cdot N(t)$) but this is equivalent to maximizing the arithmetic mean of the growth rate $\langle \mu \rangle$ as we have defined it in Eq. (1). However, this is not an exact result for finite populations, and when environmental fluctuations are strong and the extinction rate $r_{\text{ext}}$ (that is, the inverse of the mean extinction time $\tau_{\text{ext}}$) is non-negligible, as in limited-size populations, we expect to see deviations from this principle. In fact, it has been shown that in the presence of random environmental fluctuations, the mean growth rate alone is not a good predictor of the extinction time or the invasibility of a community by a competing species, as it disregards the effects of population fluctuations [15–17].

To date, many studies have explored the probability of fixation of single mutations in fluctuating environments [4, 8, 11, 18, 19], but not within the context of bet-hedging. In contrast, while other studies have extensively characterized the properties and trade-offs of different bet-hedging strategies [3, 14, 20–22], they have not looked into what the dynamical process of evolution would actually select. For example, [5] reports on evolutionary (semi-deterministic) simulations in changing environments, but did not study in detail the effect of demographic constraints on the evolutionary dynamics.

This gap in the literature is particularly relevant because, as reported in [22, 23] and as we will show in Sec. III, there exist strategies in this context that achieve a noticeable reduction in fluctuations with almost no cost to the average growth rate. The existence of these strategies suggests that the traditional focus on maximization of the mean growth rate alone, as established by modern coexistence theory [24, 25], may be insufficient to predict evolutionary outcomes in finite populations where fluctuations may play a significant role. Exactly which of these bet-hedging strategies is favored by the actual process of natural selection remains an open question, and it is the one we aim to address with this work.

The rest of this paper is organized as follows. In Sec. II, we present the details of the agent-based model developed to study this problem. In Sec. III, we characterize the system's dynamics in the absence of mutations. In Sec. IV, we introduce mutations to analyze the evolutionary dynamics and identify the selected strategies. To shed light on these results, we propose a heuristic approximation of the distribution of strategies during the evolutionary process in Sec. V. We then demonstrate the robustness of our findings by exploring a different mutation mechanism in Sec. VI, and a different set of parameters in Sec. VII. Finally, in Sec. VIII, we discuss the physical and biological implications of our work and present some concluding remarks and outlooks.

## II. THE MODEL: A MARKOV PROCESS

To address the question outlined in the introduction, we introduce the simple agent-based model illustrated in Fig. 1. The state of the system consists of only two components: the *environment* and the *population*.

The *environment* is a discrete variable with two possible values $e \in \{1, 2\}$, which evolves according to a continuous-time Markov chain with transition rates $\omega_t(e, e')$ between states $e$ and $e'$. This framework could easily be generalized to the case of $n_{\text{env}}$ different environments.

The *population* consists of $N$ agents, each characterized by two traits: a *phenotype* with two possible values $\phi \in \{A, B\}$ (which, again, could easily be generalized to more than two phenotypes), and a probability distribution over these possible phenotypes, $s(\phi)$. We will refer to this distribution as the agent's *phenotypic strategy*. The phenotype will determine the response of the agent to the environment, since the population will follow a stochastic birth-death process in which the birth and death rates ($\omega_b(e, \phi)$ and $\omega_d(e, \phi)$ respectively) depend explicitly on both the current environment and the phenotype of each agent.

On the other hand, Fig. 1(c), the phenotypic strategy will regulate the phenotypic composition of the population: after agent replication, the offspring will exhibit the phenotype $A$ with probability $s(A)$ and the phenotype $B$ with probability $s(B)$ (since we restrict ourselves to two phenotypes, the strategy is fully specified by a single parameter, $s(A)$, as $s(B) = 1 - s(A)$). Although agents' phenotypes are determined at birth and remain fixed, this mechanism enables phenotypic diversity to emerge and be maintained *across generations* (provided that it is favorable to the population). Additionally, when an agent replicates, its offspring will generally inherit the phenotypic strategy of the parent $s$, but with probability $p_{\text{mut}}$ a mutation will occur, leading to a new random strategy $s_{\text{mut}}$. This will enable the evolutionary exploration of the space of possible phenotypic strategies, which constitute the heritable traits subject to selection, and therefore play the role of the *genotype*. For simplicity, mutations in this model are assumed to be abrupt, although this may not be the most common situation in biological systems. In Sec. VI, we extend the model by introducing an additional parameter to allow for incremental mutations, and show that this modification leads to qualitatively similar results, with only minor quantitative differences.

The mechanism we use to introduce bet-hedging differs from that adopted in many previous works [3, 5, 14, 21, 22, 26] which rely on stochastic phenotypic switching. In contrast, as stated above, in our approach phenotype is fixed at birth (as in [27]). With this simplification, the strategy is specified with a single variable, facilitating the interpretation of the selected strategies while providing a generic and biologically relevant framework for modeling bet-hedging.

Finally, as we pointed out in the introduction, we expect to see deviations from the already known principles of evolution when the population size does not grow indefinitely. Therefore, we will limit the population size to its initial value $N_{\text{ini}}$: whenever replication would cause the population to exceed this value, a randomly chosen agent is simultaneously removed. This mechanism mimics continuous-culture evolution experiments [12], and can be interpreted as imposing an effective ecological carrying capacity while preserving exponential growth dynamics.

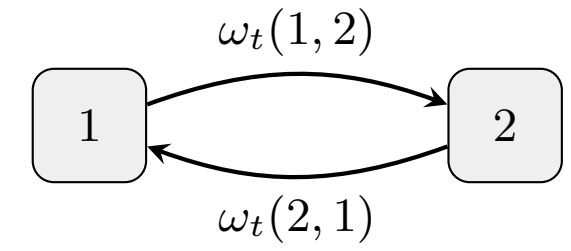

(a) Environment dynamics

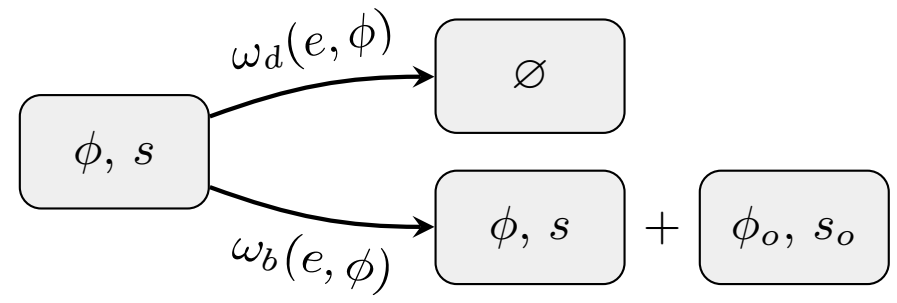

(b) Birth-death process

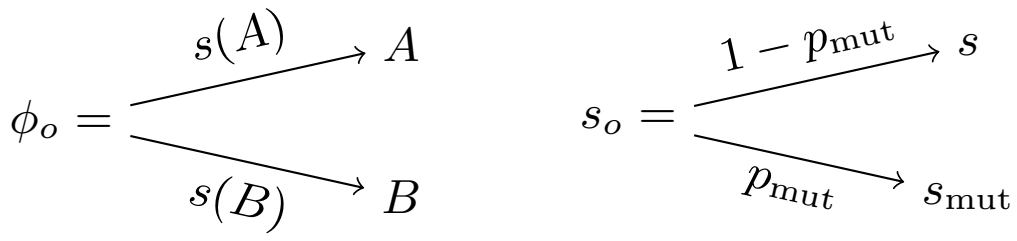

(c) Phenotype and strategy of the offspring

FIG. 1. Schematic diagram of our model: the *environment* variable follows a simple Markov process between two states (a), the population follows a *birth-death process* with rates that depend on the environment and the phenotypes of the agents (b) and finally, the phenotype of the offspring will be determined by the strategy of the parent and the offspring will inherit that same strategy unless a mutation occurs (c).

Despite the minimalism of the model, it is analytically intractable. Therefore, we will perform simulations using the Gillespie algorithm [28]. More details about the simulations and the algorithm can be found in Appendix . Nevertheless, as we will see in the next sections, the model is simple enough that its behavior can be readily analyzed, and at the same time it contains all the necessary elements to provide insight into the questions that motivate this work.

## III. BEHAVIOR OF THE MODEL WITHOUT EVOLUTION

We begin by simulating the model without evolution, i.e. with $p_{\text{mut}} = 0$, setting the initial population size to $N_{\text{ini}} = 100$ and assuming that all agents initially share the same strategy, $s_{\text{ini}}$. This kind of simulations (which we will call `fixed`) allow us to study the properties of individual strategies in isolation and, in turn, to better understand the evolutionary outcomes that follow.

To proceed, we must first choose a specific set of model parameters. We start with the *Persister–Resistant* (PR) parameter set shown in Table I. A completely different parameter set is considered in Sec. VII, yielding qualitatively similar results. The PR parameter set reflects a common biological scenario involving two distinct phenotypes: a persister and a resistant phenotype [5, 14]. Specifically, the parameters are defined as follows: both environments are equally likely, with an environmental transition rate of 1.0. Environment 1 represents a "favorable" environment in which phenotype $A$ (the resistant type) replicates at a rate of 1.0, while phenotype B (the persister type) replicates at a rate of 0.2, neither incurring death. Conversely, environment 2 represents "harsh" conditions in which phenotype $A$ dies at a rate of 1.0 and phenotype $B$ at a rate of 0.1, with neither of them capable of replication.

| symbol | name | value |
|---|---|---|
| $\omega_t(e,e')$ | transition rates | $\begin{pmatrix} -1.0 & 1.0 \\ 1.0 & -1.0 \end{pmatrix}$ |
| $\omega_b(e,\phi)$ | birth rates | $\begin{pmatrix} 1.0 & 0.2 \\ 0.0 & 0.0 \end{pmatrix}$ |
| $\omega_d(e,\phi)$ | death rates | $\begin{pmatrix} 0.0 & 0.0 \\ 1.0 & 0.1 \end{pmatrix}$ |

TABLE I. Set of parameters used in the simulations of Secs. III to VI, which we will refer to as the *Persister–Resistant* (PR) parameter set.

Figure 2 shows the extinction rate $r_{\text{ext}}$ versus the mean growth rate $\langle\mu\rangle$ in simulations with the PR parameter set. Here, we restrict our attention to the `fixed` simulations, represented by the blue curve, with strategies ranging from $s(A) = 0$ (lower end of the curve) to $s(A) = 1$ (upper end). This curve, known as a *Pareto front*, represents the set of optimal trade-offs: any increase in average growth rate necessarily comes at the cost of a higher extinction risk and vice versa. In more general settings with more than two phenotypes, strategies would span a higher-dimensional space, and the Pareto front would correspond to the boundary (also known as the efficient frontier) of the accessible region.

The first important observation is that the maximum of $\langle\mu\rangle$ is achieved neither at $s(A) = 0$ nor at $s(A) = 1$, but rather at an intermediate strategy, indicating that in scenarios such as this one, phenotypic diversity (that is, maintaining a mixture of both phenotypes) is the strategy which maximizes the average growth rate, a result already reported several times [3, 22, 23, 27]. As we will see, this is precisely the strategy reached by evolution at large $N_{\text{ini}}$. This behavior does not arise in all cases, but only when environmental fluctuations make each phenotype individually less favorable in the long term than their combination. Secondly, we observe that the growth-maximizing strategy differs from the extinction-minimizing one (again, this holds true for other parameter sets, as discussed in Sec. VII). In particular, $\langle\mu\rangle$ is maximized at $s(A) \sim 0.8$, while $r_{\text{ext}}$ is minimized at $s(A) \sim 0.0$. As we show in Sec. V, this trade-off (and more precisely the different scaling of these two quantities with the population size) plays a key role in shaping the result of evolution. Note that all figures throughout this work include error bars or error bands; however, in many cases, they are smaller than the point symbols and are therefore barely visible.

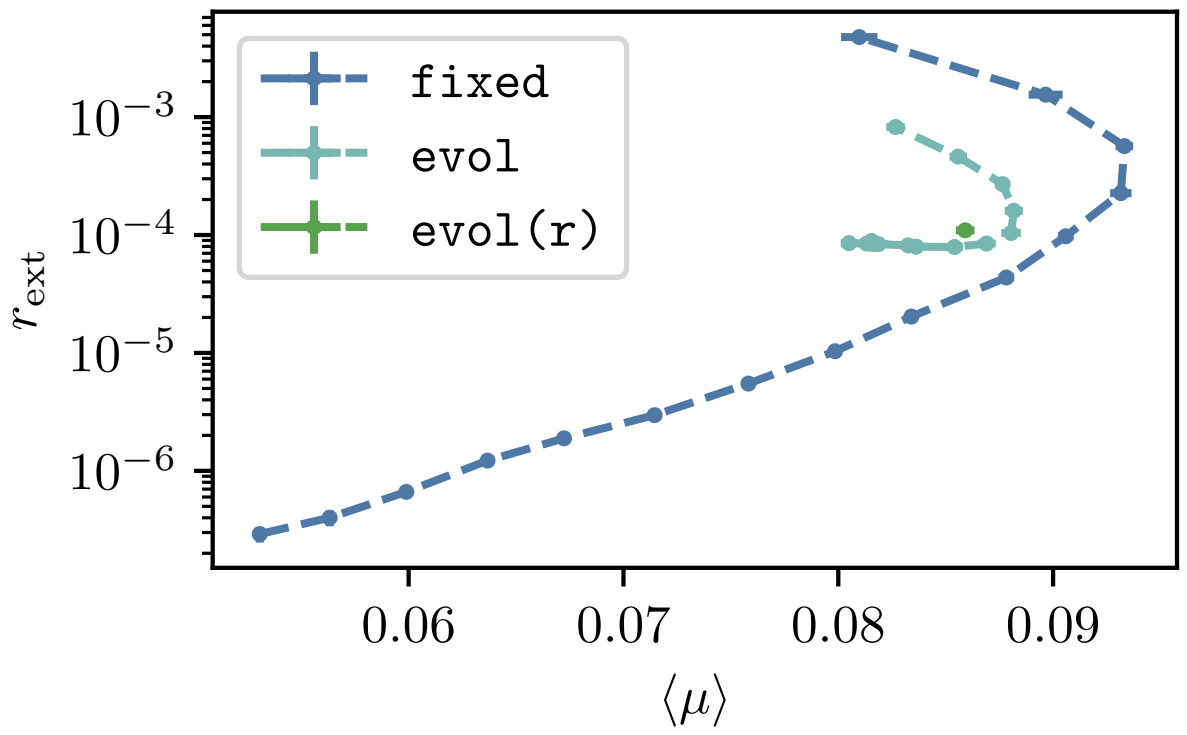

FIG. 2. Extinction rate $r_{\text{ext}}$ versus mean growth rate $\langle\mu\rangle$. Results are shown for `fixed`, `evol` and `evol(r)` simulations, across a range of initial strategies from $s(A) = 0$ (lower end of the curve) to $s(A) = 1$ (upper end). These values define a *Pareto front* capturing the trade-off between growth and survival.

It is important to clarify that many of the quantities analyzed throughout this work (like $\langle\mu\rangle$ in Fig. 2) represent time averages and their estimation requires good statistics. However, since populations eventually undergo extinction under these stochastic dynamics (often on timescales too short to gather sufficient data from a single run), we concatenate multiple independent realizations into a single longer trajectory. All time averages and distributions are subsequently computed across these extended, joint trajectories.

Rather than exploring in detail how these results depend on the specific choice of parameters (a topic extensively studied in previous work [3, 5, 22]) the next section focuses on the evolutionary dynamics and the strategies ultimately selected under these conditions.

## IV. EVOLUTIONARY DYNAMICS

We now enable evolution by setting $p_{\text{mut}} = 0.001$ (while keeping $N_{\text{ini}} = 100$), allowing strategies to mutate and compete. We consider two types of simulations: in `evol` simulations, the population is initially composed of agents with the same initial strategy $s(A)_{\text{ini}}$, while in `evol(r)` simulations, the initial strategy of each agent is completely random.

First, to examine the global effect of this evolutionary process, in Fig. 2 we also display the values of $r_{\text{ext}}$ and $\langle\mu\rangle$ obtained when evolution is enabled (both for `evol` and `evol(r)` simulations). This allows us to directly com-pare the resulting growth and extinction rates against the fixed-strategy baseline. We find that the system converges (since the values of $r_{\rm ext}$ and $\langle\mu\rangle$ concentrate), with only a moderate dependence on the initial conditions, toward a point with a lower growth rate than the growth-maximizing strategy, but also with a reduced extinction rate.

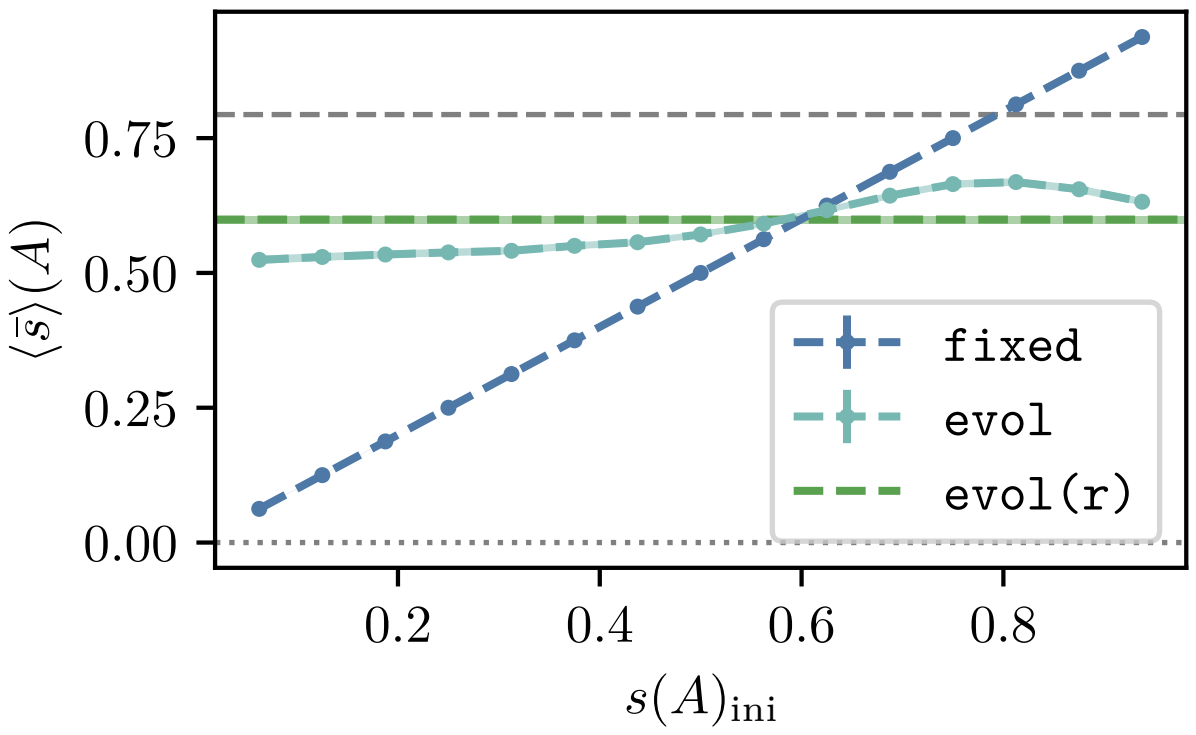

FIG. 3. Time average of the population-averaged strategy, $\langle\bar{s}\rangle(A)$, as a function of the initial strategy $s(A)_{\rm ini}$ for the PR parameter set. Results are shown for **fixed**, **evol** and **evol(r)** simulations (represented as a horizontal line, as they lack a single defined initial strategy). The dashed and dotted lines indicate the strategies that maximize mean growth and minimize extinction, respectively, with the evolved strategy lying between them.

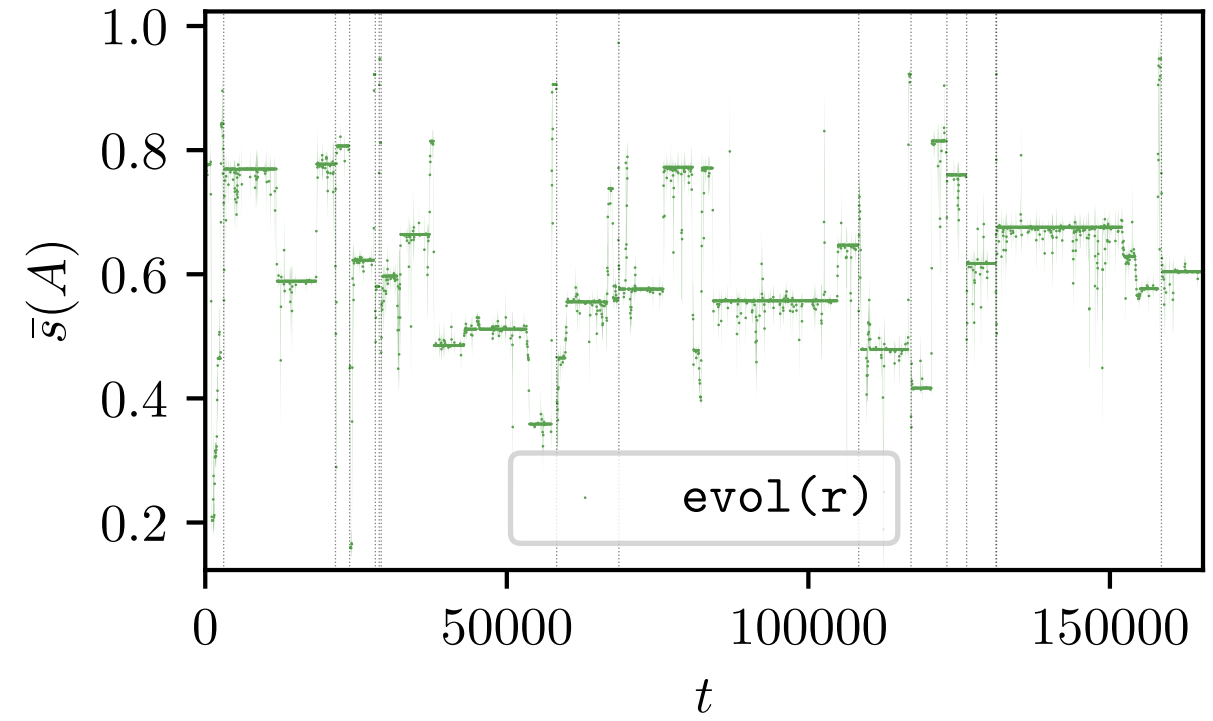

FIG. 4. Time series of the population-averaged strategy over multiple concatenated simulations (extinction events are marked by vertical dotted lines). The error band displays the strategy standard deviation within the population at each instant, $\sigma_s$, which remains remarkably narrow. Notably, before extinction is reached, advantageous mutations frequently emerge and fix, abruptly shifting the average strategy and driving a broad exploration of the strategy space.

To better characterize the result of the evolutionary process, we also compute the population-averaged strategy $\bar{s}(A)$ and analyze its time average, denoted as $\langle\bar{s}\rangle(A)$. In Fig. 3 we show the averaged strategy $\langle\bar{s}\rangle(A)$ as a function of the initial strategy $s(A)_{\rm ini}$. For **fixed** simulations the initial strategy is preserved, producing a diagonal line. In contrast, for evolutionary simulations, the system consistently converges toward a well-defined stationary value, exhibiting only a weak dependence on the initial condition. Moreover, we observe that the average evolved strategy lies between the growth-maximizing and extinction-minimizing strategies identified in the fixed-strategy simulations. As we will see in Sec. VII, this behavior is not specific to this parameter set. This shows that evolution in finite populations does not optimize instantaneous growth alone but instead selects strategies that represent a compromise between growth and extinction avoidance.

A key observation is that the instantaneous dispersion of strategies $\sigma_s$ (defined as the standard deviation of strategies present in the population at a given time) remains small. This dispersion is represented in Fig. 3 as an error band, which is barely visible. At any given time, the population is, therefore, effectively characterized by a single dominant strategy. Over longer times, however, mutation-driven exploration causes this dominant strategy to drift through the space of possible strategies, as can be seen in Fig. 4. There we show the time evolution of the population-averaged strategy in the **evol(r)** simulations. Extinction events are clearly identified by vertical dotted lines. Notably, during evolution, abrupt changes in the average strategy are observed: beneficial mutations repeatedly arise and fix, thus promoting extensive exploration of the strategy space.

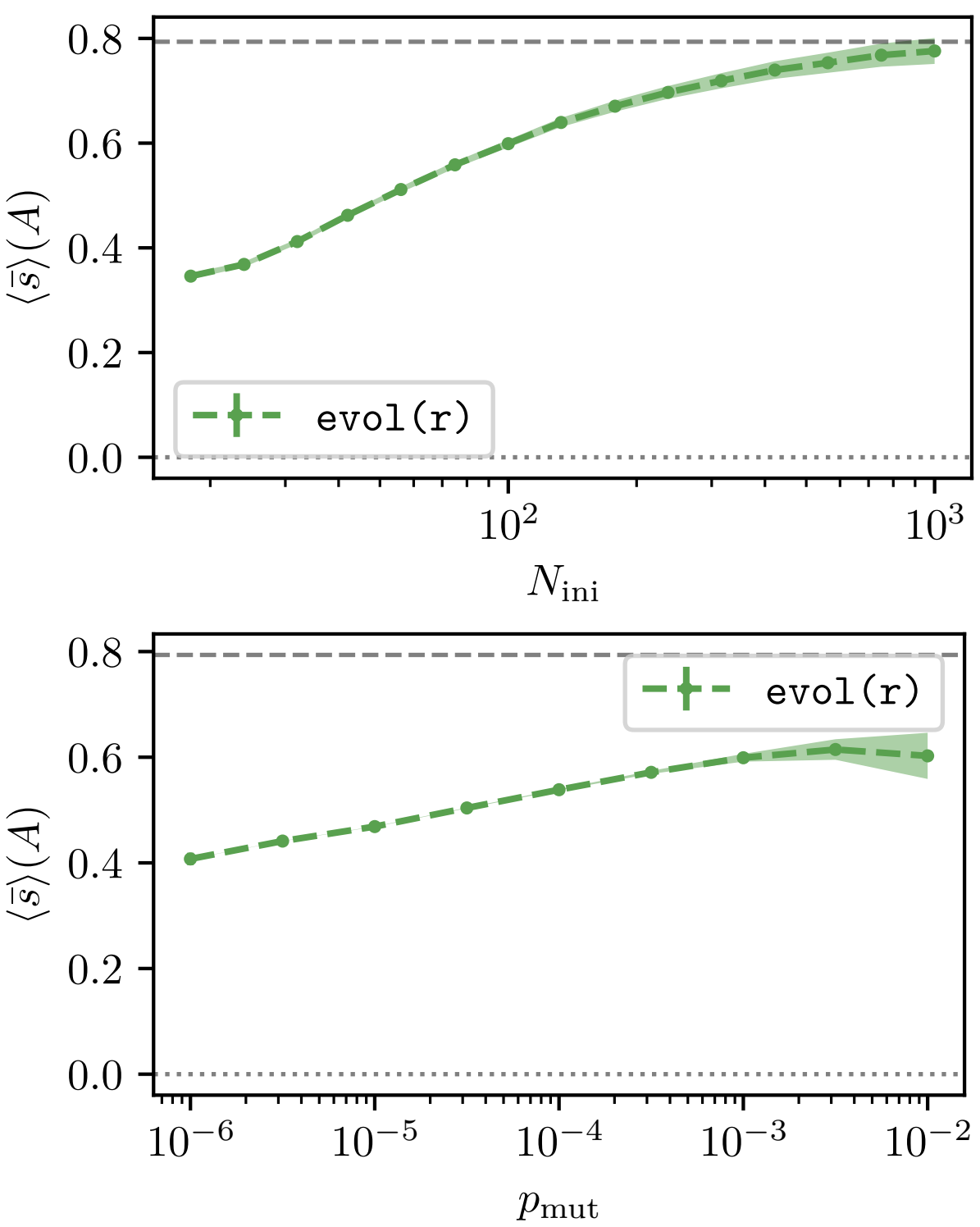

FIG. 5. Dependence of the average evolved strategy on the initial population size $N_{\rm ini}$ (top) and the mutation probability $p_{\rm mut}$ (bottom). Increasing $N_{\rm ini}$ drives the system toward the growth-maximizing strategy, while smaller populations favor safer strategies with lower extinction rates. The dashed and dotted lines indicate the growth-maximizing and extinction-minimizing strategies, respectively, and the error band displays the dispersion of strategies $\sigma_s$.

Finally, in Fig. 5 we explore how the evolutionary outcome (namely, the average strategy $\langle \bar{s} \rangle(A)$) depends on the size of the population $N_{\text{ini}}$ and the mutation probability $p_{\text{mut}}$. We first observe a strong dependence on $N_{\text{ini}}$. As anticipated, when $N_{\text{ini}}$ increases, $\langle \bar{s} \rangle(A)$ approaches the growth-maximizing value identified in Sec. III. Only for small populations does evolution favor safer strategies with lower extinction rates. In contrast, the dependence on the mutation probability is relatively weak, as $p_{\text{mut}}$ mainly controls the exploration of the strategy space (in Sec. VI we further examine the effect of changing the characteristics of mutations). For very large values of $p_{\text{mut}}$, the strategy standard deviation $\sigma_s$ increases, indicating that the population no longer shares a common strategy at a given time. In fact, in the limit $p_{\text{mut}} \to 1$, the population behaves as if $s(A) = 0.5$, since the strategies change randomly with each replication. Conversely, in the limit $p_{\text{mut}} \to 0$, mutations are absent and the result is determined solely by selection among the strategies present in the initial condition.

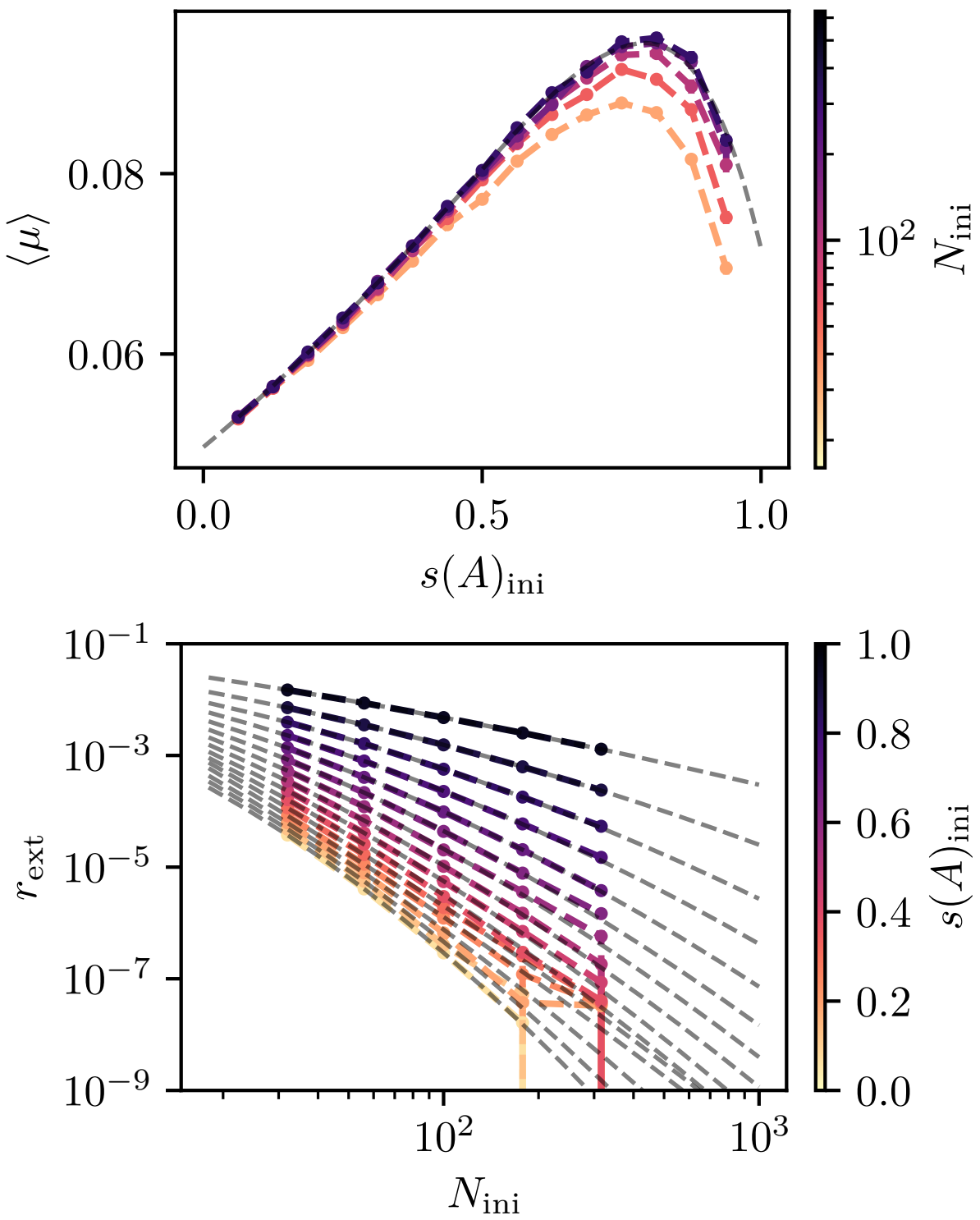

FIG. 6. Results from `fixed` simulations used to construct the heuristic approximation. The mean growth rate (top) depends only weakly on $N_{\text{ini}}$, whereas the extinction rate (bottom) follows an approximate power-law scaling. Dashed lines indicate spline interpolations of the simulation data.

## V. HEURISTIC INTERPRETATION OF THE EVOLUTIONARY OUTCOME

To understand the result of the evolutionary dynamics, a key observation must be made regarding Fig. 2: the points representing the evolutionary simulations do not lie on, nor do they converge toward, the Pareto front defined by strictly fixed strategies. This indicates that the evolutionary process cannot be associated with any single static fixed strategy.

However, as previously discussed, along the evolutionary process the population is effectively characterized by a single dominant strategy at any given moment, which slowly shifts over time due to mutations and reinitializations. Over long timescales, the system can therefore be interpreted as a temporal concatenation of fixed-strategy dynamics with varying values of the population-averaged strategy $\bar{s}$. Consequently, we should be able to approximate the average growth and extinction rates in evolutionary simulations by weighting the properties of fixed strategies by the actual distribution of strategies explored during the evolutionary process:

$$\langle \mu \rangle \approx \int \langle \mu \rangle(\bar{s}) p(\bar{s}) d\bar{s}, \quad r_{\text{ext}} \approx \int r_{\text{ext}}(\bar{s}) p(\bar{s}) d\bar{s}, \qquad (2)$$

where $\langle \mu \rangle(\bar{s})$ and $r_{\text{ext}}(\bar{s})$ represent the mean growth rate and extinction rate of a strictly *fixed* strategy with value $\bar{s}$ (the blue curve in Fig. 2), while $p(\bar{s})$ is the probability distribution of the population-averaged strategy $\bar{s}$ recorded over time during the active evolutionary simulations. We have verified that this approximation holds with high accuracy across a wide range of parameters [29]. This fact motivates the search for a heuristic approximation of the distribution $p(\bar{s})$, obtained solely from properties of fixed strategies, that could allow us to predict the evolutionary outcome $\langle \bar{s} \rangle(A)$ as a function of $N_{\text{ini}}$ and $p_{\text{mut}}$.

To construct this approximation, we assume first that the probability of a given strategy being selected during the initial competitive phase following reinitialization is proportional to $\exp(N_{\text{ini}} \cdot \langle \mu \rangle(\bar{s}))$. This is because the differences in the average growth rate will be exponentially amplified over the duration of this competitive phase and the time required to resolve this competition will scale with the number of competing strategies (as a first approximation, we will assume that this duration will be directly proportional to $N_{\text{ini}}$, which justifies its presence in the exponent). However, to obtain the long-term distribution of strategies, this selection probability must be weighted by the expected survival time before the population goes extinct or is replaced by a mutant. This leads us to divide the exponential term by the sum of two distinct rates: the population extinction rate $r_{\text{ext}}(N_{\text{ini}}, \bar{s})$, and the mutation-driven replacement rate. While an exact derivation of the latter is generally prohibitive, it acts as a minor correction where an order-of-magnitude estimate is sufficient. This replacement rate can be estimated as the total mutation emergence rate, $N_{\mathrm{ini}} \cdot p_{\mathrm{mut}} \cdot \langle\mu_b\rangle(\bar{s})$ (where $\langle\mu_b\rangle$ is the average birth rate, excluding deaths), multiplied by the fixation probability of the new mutant. Although exactly calculating this probability is very challenging [4, 8, 11], it can be approximated as $1/N_{\mathrm{ini}}$, causing $N_{\mathrm{ini}}$ to cancel out. Together, these terms yield the following heuristic form for the distribution of strategies:

$$p(\bar{s}) \propto \frac{\exp(N_{\mathrm{ini}} \cdot \langle\mu\rangle(\bar{s}))}{r_{\mathrm{ext}}(N_{\mathrm{ini}}, \bar{s}) + p_{\mathrm{mut}} \cdot \langle\mu_b\rangle(\bar{s})}. \quad (3)$$

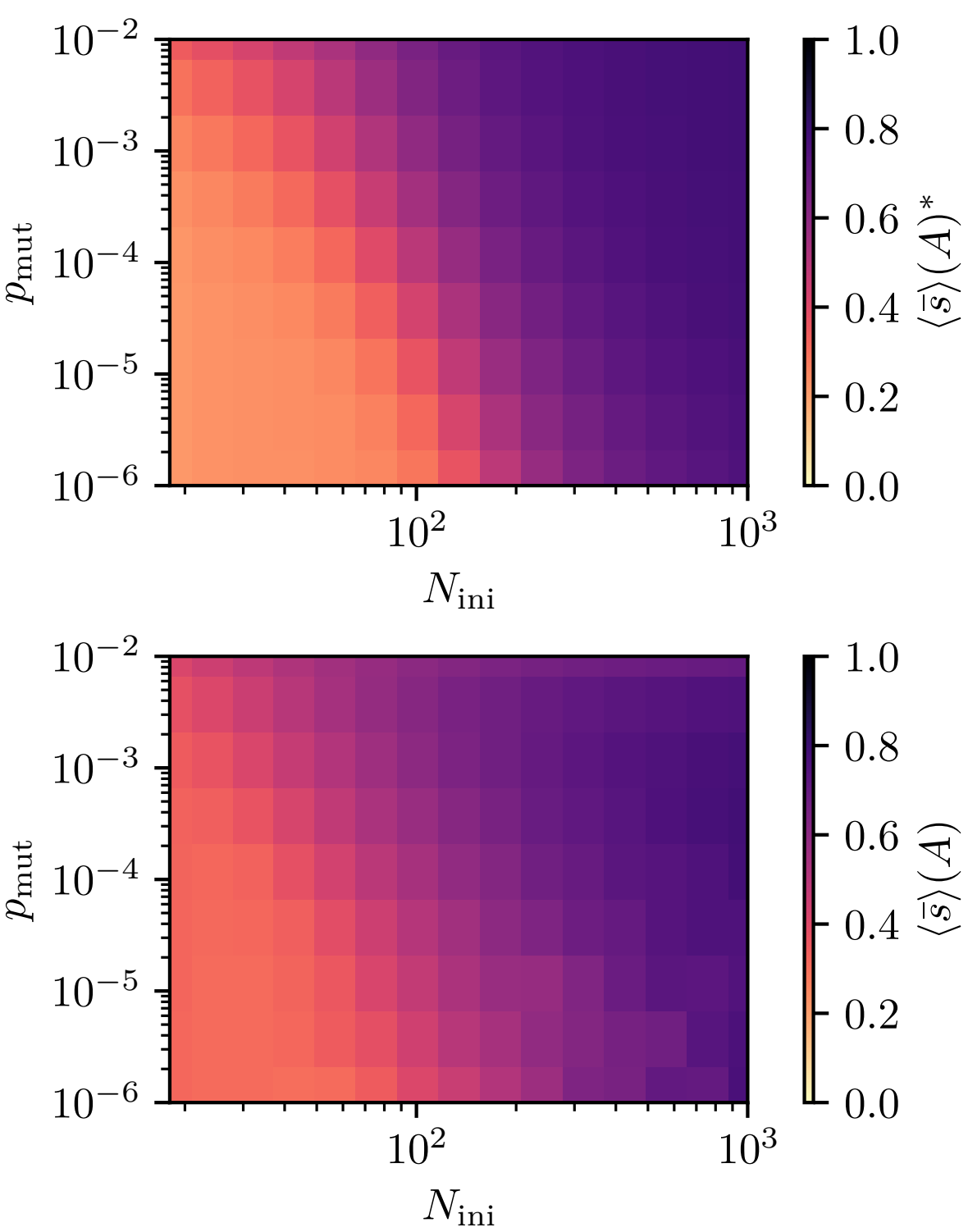

FIG. 7. Expected average strategy from the heuristic distribution (top) and observed average strategy from `evol(r)` simulations (bottom) for different values of $N_{\mathrm{ini}}$ and $p_{\mathrm{mut}}$.

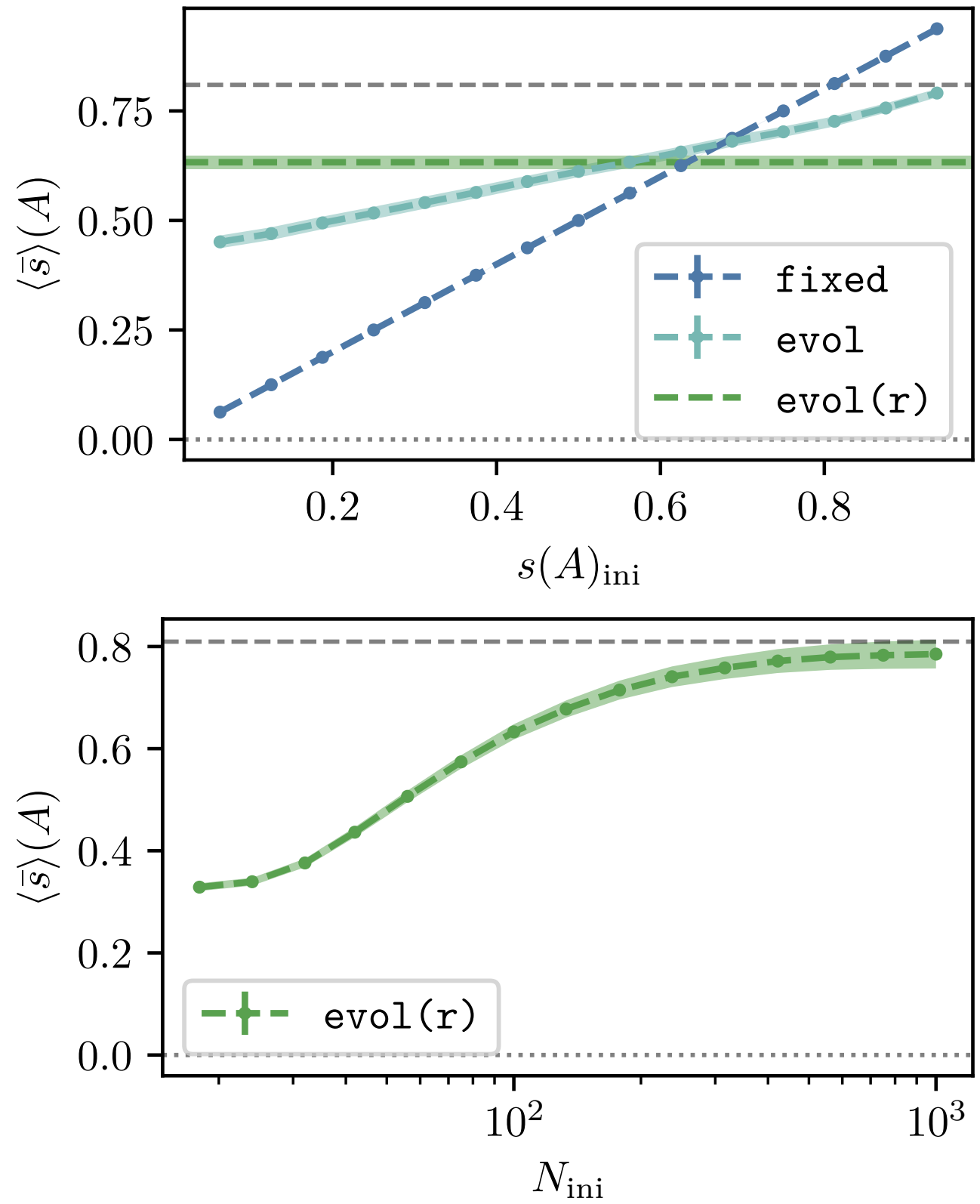

FIG. 8. Dependence of the evolved strategy on initial conditions with $N_{\mathrm{ini}} = 100$ (top) and on population size (bottom) for the PR parameter set with incremental mutations. The dashed and dotted lines indicate the strategies that maximize mean growth and minimize extinction, respectively, and the error band the dispersion of strategies $\sigma_s$.

To evaluate $\langle\mu\rangle$, $\langle\mu_b\rangle$ and $r_{\mathrm{ext}}$ for arbitrary values of $\bar{s}$ and $N_{\mathrm{ini}}$ we use spline interpolations of the numerical data obtained from `fixed` simulations shown in Fig. 6.

To understand how the distribution depends on the model parameters we examine how $\langle\mu\rangle$ and $r_{\mathrm{ext}}$ depend on the population size; see Fig. 6. We find that the mean growth rate depends only weakly on $N_{\mathrm{ini}}$, except for very small populations, making it primarily a function of $s$. In contrast, the extinction rate exhibits a strong dependence, decreasing approximately as a power law, $r_{\mathrm{ext}} \sim N_{\mathrm{ini}}^{-\alpha}$. This scaling arises because extinctions are primarily driven by unfavorable environmental episodes. In an unfavorable environment we will have $N(t) \sim N_{\mathrm{ini}} e^{-\lambda t}$ with some effective decay rate $\lambda$, meaning that extinction requires an adverse episode of duration $T \sim \ln(N_{\mathrm{ini}})/\lambda$. Since environmental durations are exponentially distributed, the extinction rate scales as $e^{-T}$, which leads to the power-law dependence $r_{\mathrm{ext}} \sim N_{\mathrm{ini}}^{-1/\lambda}$. Consequently, in Eq. (3), the growth contribution enters exponentially through the numerator, whereas extinction risk enters algebraically in the denominator. This scaling mismatch is the main reason behind the transition: extinction risk dominates at small population sizes, favoring safer strategies, while growth maximization inevitably prevails in the large $N_{\mathrm{ini}}$ limit.

Figure 7 compares the expected average strategy calculated from the heuristic distribution (top) with the actual average observed in evolutionary simulations (bottom). While a detailed comparison of the full probability distributions is deferred to Sec. VIII, focusing on their mean values reveals good qualitative agreement in a wide range of $N_{\mathrm{ini}}$ and $p_{\mathrm{mut}}$ values. Crucially, the heuristic successfully captures the transition from growth-maximizing to safer strategies as $N_{\mathrm{ini}}$ decreases, as well as how this transition is modulated by $p_{\mathrm{mut}}$.

## VI. INCREMENTAL MUTATIONS

As anticipated when introducing the model, biological mutation mechanisms can take various forms, and the most adequate one to model them often depends on the specific system under study. To demonstrate that our core insights and conclusions are robust to these microscopic details, we extend the model by considering an incremental mutation mechanism rather than completely random, abrupt changes. Specifically, when a mutation occurs, a Gaussian random variable with standard deviation $\sigma_{\text{mut}}$ is added to each component of the parent's strategy, which is subsequently normalized.

To keep the effect of mutations comparable to that of the main model, we choose $\sigma_{\text{mut}} = 0.1$ and $p_{\text{mut}} = 0.01$, ensuring that their product matches the mutation probability previously used. As shown in Fig. 8, the qualitative behavior under incremental mutations is remarkably consistent with the main model: evolution still leads to intermediate strategies, preserving the transition from growth maximization at large $N_{\text{ini}}$ to more resilient strategies at smaller population sizes. However, we observe a slightly stronger dependence on initial conditions compared to Fig. 3. This is likely because mutations are now continuous, which creates smoother trajectories that make escaping unfavorable regions of the strategy space more difficult.

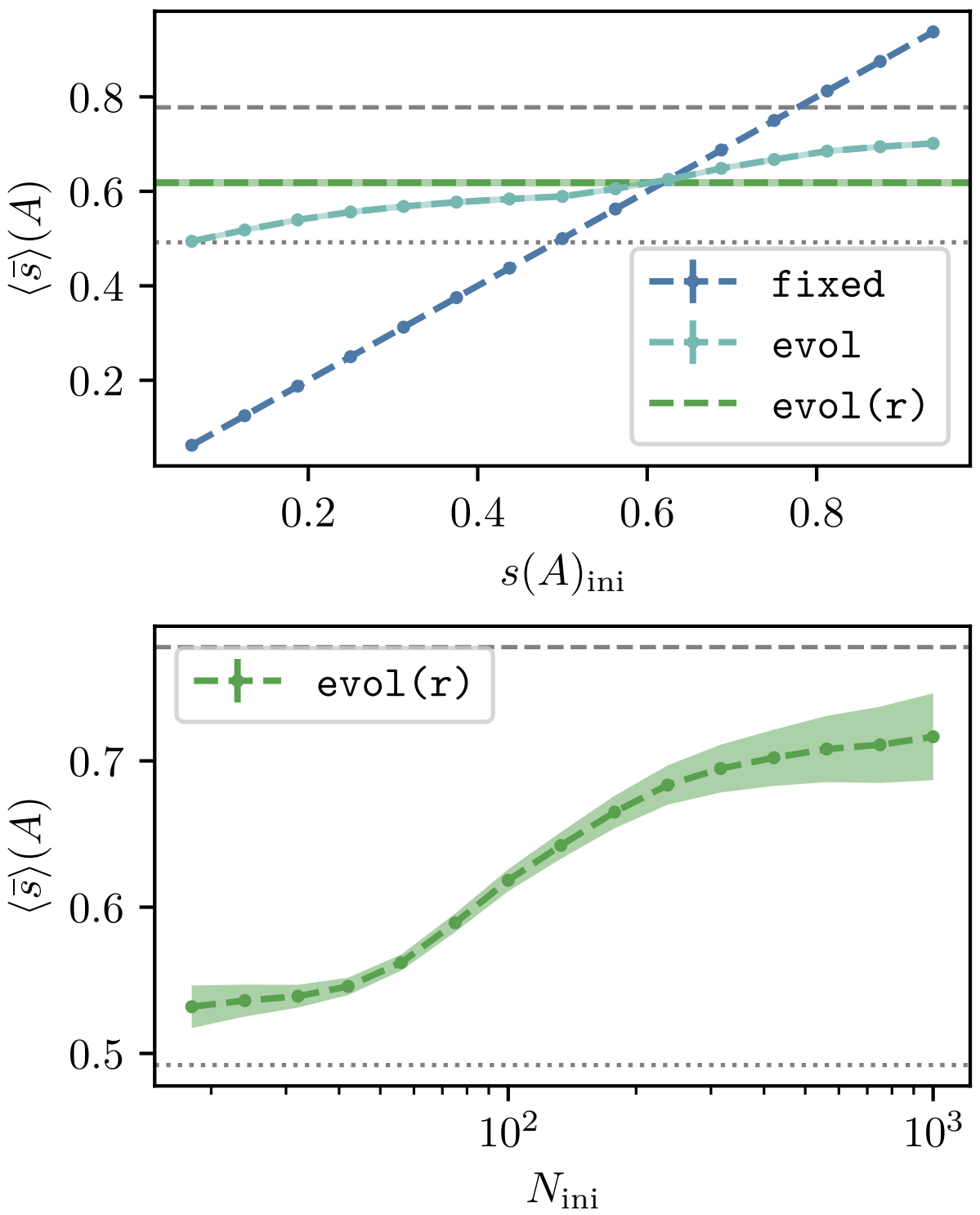

FIG. 9. Dependence of the evolved strategy on initial conditions with $N_{\text{ini}} = 100$ (top) and on population size (bottom) for the SA parameter set. The same qualitative trends as in the PR parameter set are observed. The dashed and dotted lines indicate the strategies that maximize mean growth and minimize extinction, respectively, and the error band the dispersion of strategies $\sigma_s$.

## VII. A SECOND SET OF PARAMETERS

To further assess the generality of our findings, we analyze a qualitatively different parameter set, which we denote as the Symmetric–Adaptation (SA) set. It is defined by the same environment transition rates $\omega_t(e, e')$ as in the PR parameter set and the following birth and death rates:

$$\omega_b(e, \phi) = \begin{pmatrix} 1.2 & 0.0 \\ 0.0 & 0.8 \end{pmatrix}, \quad \omega_d(e, \phi) = \begin{pmatrix} 0.0 & 1.0 \\ 1.0 & 0.0 \end{pmatrix}. \qquad (4)$$

In this scenario, each phenotype is highly specialized for one specific environment and performs poorly in the other, representing a fundamentally different ecological setup. Although with different parameter values, this symmetric setup has been analyzed before [3, 20] and could for instance represent different growth rates of the two cell types (induced or uninduced for lactose) in the presence of glucose or lactose in the classic diauxie experiment, or the expression (absence) of pili in presence (absence) of urine flow in bacteria [30]. As in Sec. IV, we use the original mutation mechanism with $p_{\text{mut}} = 0.001$.

In Fig. 9 we show the evolved strategy for different initial conditions and population sizes. Again, the SA parameter set also exhibits a trade-off between growth and extinction, although the optimal strategies are much closer to each other than in the PR case. Specifically, the growth-maximizing strategy is located at $s(A) \sim 0.8$, whereas the extinction-minimizing strategy is found at $s(A) \sim 0.5$. Despite this proximity our main conclusions still hold: evolutionary trajectories converge to a well-defined strategy largely independent of initial conditions, lying strictly between both optima. The population size $N_{\text{ini}}$ again acts as the key control parameter: small populations favor safer strategies (getting remarkably close to the extinction-minimizing one) while larger populations favor growth maximization. Notably, with these parameters, substantially larger values of $N_{\text{ini}}$ are required to approach the growth-maximizing strategy compared to the PR parameter set.

Finally, the heuristic approximation still successfully captures this qualitative transition between regimes. However, in this case quantitative discrepancies are more pronounced (see details in Sec. VIII). In particular, approaching the growth-maximizing strategy requires significantly larger $N_{\text{ini}}$ values than predicted by the heuristic approximation.

## VIII. DISCUSSION AND OUTLOOK

Our results provide a clear and minimal illustration of how the result of evolution in finite populations cannot be inferred solely from average instantaneous growth rates. The long-term dominant strategy is shaped not only by the mean behavior of the population, but also by fluctuations, particularly those that drive extinction. In fluctuating environments with limited carrying capacity, extinction becomes inevitable over long timescales and thus acts as a genuine selective pressure. In the large-$N_{\text{ini}}$ limit, extinction risk becomes negligible, and evolution selects the strategy that maximizes $\langle\mu\rangle$ (which, as expected in fluctuating environments, can be a bet-hedging strategy where $s(A)$ is neither 0 nor 1). For smaller populations, however, extinction risk reshapes the adaptive landscape, biasing evolution toward safer, lower-growth strategies. Thus, demographic constraints fundamentally alter the nature of evolutionary optimization.

The heuristic approximation introduced here captures the qualitative competition between growth and extinction, clarifying the scaling mechanisms underlying the transition from growth-maximizing to resilient strategies. Quantitative discrepancies remain, highlighting that evolutionary dynamics cannot be fully understood solely through the properties of fixed strategies. Refining this heuristic approximation and, more broadly, deriving further analytical results for this model represent an important avenue for future work.

Furthermore, we have verified that our core findings are robust to variations in model parameters and the specifics of the mutation mechanism. This consistency confirms that the observed transition from growth-maximizing to safer strategies is a fundamental feature of finite-population dynamics in fluctuating environments, rather than an artifact of specific modeling choices.

Ultimately, the model developed here provides a general framework for studying evolution in fluctuating environments under population constraints. Beyond theoretical modeling, designing experiments with bacterial populations to empirically test these predictions represents a promising next step. Further extensions of the model (such as expanding the number of phenotypes and environments) could also yield compelling insights, despite the added analytical complexity. Particularly promising is the inclusion of environmental sensing, allowing phenotypic strategies to adapt dynamically to environmental states or correlated cues.

## ACKNOWLEDGMENTS

We are grateful to Saúl Ares and Pablo Catalán for stimulating discussions and valuable comments on the manuscript. We wish to acknowledge financial support through Grant PID2023-147067NB-I00 funded by MCIU/AEI/10.13039/501100011033 and by ERDF/EU.

## DATA AVAILABILITY

The source code developed for this study is openly available in the repository at `https://github.com/Marco-Mendivil-Carboni/mutare`.

## APPENDIX: THE GILLESPIE ALGORITHM AND TIME SERIES ANALYSIS

To perform numerical simulations of the model developed in Sec. II, we have employed the Gillespie algorithm [28], a standard method for generating statistically exact trajectories of continuous-time Markov processes such as ours. At any given time, the state of the system is denoted by $\mathbf{X}$ and all possible "events" (the birth or death of an individual, or an environmental transition) are indexed by $j = 1, \ldots, M$, each occurring with a rate $\omega_j(\mathbf{X})$. The "total rate" is therefore

$$\Omega(\mathbf{X}) = \sum_{j=1}^{M} \omega_j(\mathbf{X}). \tag{A1}$$

The algorithm proceeds iteratively as follows. First, the time $\tau$ to the next event is drawn from an exponential distribution:

$$\tau = \frac{1}{\Omega(\mathbf{X})} \ln \frac{1}{r_1}, \quad r_1 \sim U(0,1), \tag{A2}$$

where $U(0,1)$ denotes the uniform distribution on the interval $(0,1)$. Then, the specific event $k$ that occurs is selected according to:

$$\sum_{j=1}^{k-1} \omega_j(\mathbf{X}) < r_2 \Omega(\mathbf{X}) \leq \sum_{j=1}^{k} \omega_j(\mathbf{X}), \quad r_2 \sim U(0,1). \tag{A3}$$

Finally, the system state is updated as:

$$\mathbf{X}(t+\tau) = \mathbf{X}(t) + \delta\mathbf{X}_k, \tag{A4}$$

where $\delta\mathbf{X}_k$ is the state-change vector associated with event $k$, thereby advancing the simulation to the next time step.

An important consequence of using this algorithm is that the system is naturally sampled at irregular time intervals $\Delta t_k = \tau_k$. Therefore, time averages of any observable $\mathcal{O}(\mathbf{X})$ must be computed as time-weighted averages,

$$\langle\mathcal{O}\rangle = \frac{1}{\sum_k \tau_k} \sum_k \mathcal{O}(\mathbf{X}_k)\tau_k, \tag{A5}$$

rather than simple arithmetic means. All observables considered in this work are computed this way, with the sole exception of the population growth rate fluctuations, which we denote as $\sigma_\mu$ and analyze in Sec. VIII. The issue lies in the fact that a naive calculation of the standard deviation based directly on the step-wise rates $\mu_k = \Delta N_k/(N_k\tau_k)$ yields ill-defined, diverging fluctuations due to the arbitrarily small waiting times between Gillespie steps. The most appropriate way to quantify these fluctuations is through:

$$\sigma_\mu = \left(\frac{1}{\sum_k \tau_k} \sum_k \left(\mu_k - \langle\mu\rangle\right)^2 \tau_k^2\right)^{1/2}, \tag{A6}$$

which represents the diffusion coefficient of the integrated log-population growth process ($\log N \sim \sum_k \mu_k\tau_k$) rather than the variance of an instantaneous rate.

# Supplemental Material for "*Evolutionary adaptation through bet-hedging in finite populations under fluctuating environments*"

Marco Mendívil-Carboni,[1] Juan J. Mazo,[1] and Luis Dinis[1]

[1] *Grupo Interdisciplinar de Sistemas Complejos (GISC) and Departamento de Estructura de la Materia, Física Térmica y Electrónica, Universidad Complutense de Madrid, 28040 Madrid, Spain*

This document provides several supplementary results that complement some of the insights discussed throughout the main text.

## I. AVERAGE POPULATION BIRTH RATE AND GROWTH RATE FLUCTUATIONS

We begin by displaying in Fig. S1 (left) the average birth rate $\langle \mu_b \rangle$ (which appears in the heuristic approximation in Eq. (3)) for different initial conditions with the PR parameters. We observe that $\langle \mu_b \rangle$ for fixed strategies exhibits a different behavior than the total population growth rate, increasing monotonically with $s(A)$ and reaching its maximum at $s(A) = 1$. Another observable of interest are the growth rate fluctuations $\sigma_\mu$, computed according to Eq. (A6). Such fluctuations are commonly used as a proxy for extinction risk, and, indeed, the Pareto front generated between $\sigma_\mu$ and $\langle \mu \rangle$ has been studied more frequently in the literature [1, 2]. As shown in Fig. S1 (right), where we plot $\sigma_\mu$ versus $\langle \mu \rangle$, their behavior qualitatively mirrors the extinction rate, although a simple exact analytical mapping between the two quantities remains elusive.

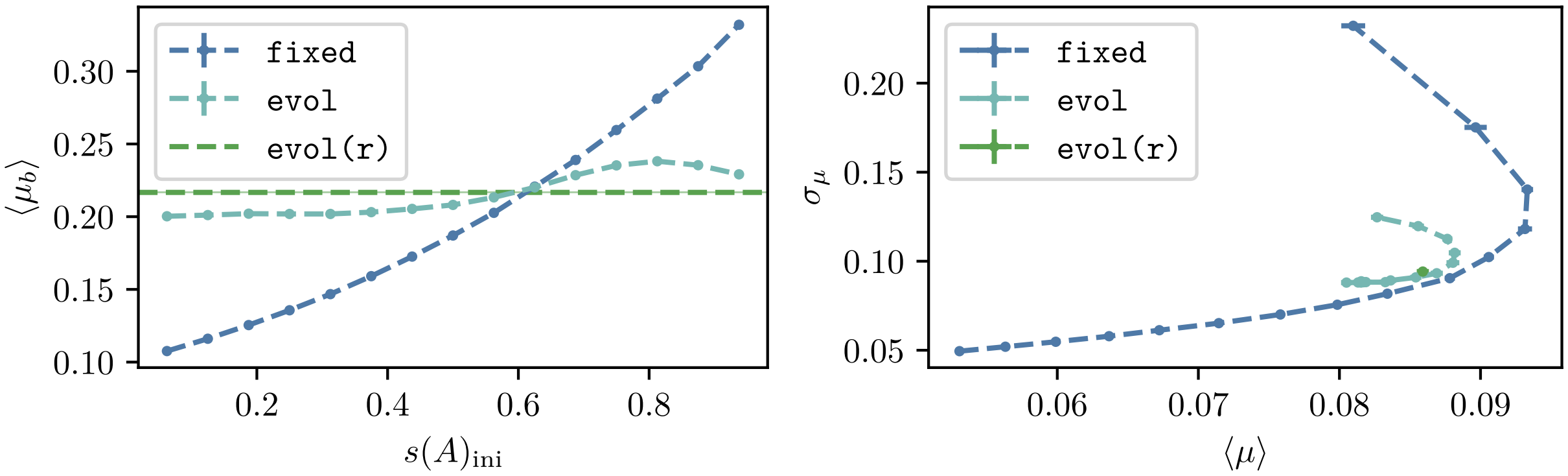

FIG. S1. Average population birth rate $\langle \mu_b \rangle$ (left) and growth rate fluctuations $\sigma_\mu$ versus mean growth rate $\langle \mu \rangle$ (right) for different initial strategies $s(A)_{\text{ini}}$ and the different kinds of simulations with the PR parameters.

## II. PHENOTYPE PROBABILITY DISTRIBUTION

We also examine the actual phenotype distribution within the population. Although dictated primarily by the strategy, this distribution does not coincide with it due to phenotype-dependent lifetimes. In Fig. S2 (left), we plot the actual probability of the resistant phenotype $A$, which increases monotonically with $s(A)$. This observable allows for a useful consistency check: the theoretical $p(A)$ can be analytically derived for static environments and fixed strategies from the dominant eigenvector of the matrix

$$\begin{pmatrix} s(A)\omega_b(e,A) - \omega_d(e,A) & s(A)\omega_b(e,B) \\ s(B)\omega_b(e,A) & s(B)\omega_b(e,B) - \omega_d(e,B) \end{pmatrix}, \tag{S1}$$

which governs the deterministic population dynamics in a fixed environment $e$. The two dashed lines in this figure represent the expected analytical values of $p(A)$ for different fixed strategies in each of the environments. We observe that in fluctuating environments $\langle p(A) \rangle$ systematically lies between these two limits.

Additionally, we evaluate the distribution of the population size $N$, as shown in Fig. S2 (right). All profiles reveal a pronounced peak at the maximum allowed population capacity, indicating that the system spends a significant fraction of time near its ceiling when environmental conditions are favorable. However, a second, much broader local maximum emerges at lower population sizes, which is a direct consequence of periods under the unfavorable environment. The position and weight of this second peak strongly correlate with the extinction risk: lower values of $N$ at this local maximum directly correspond to a higher extinction rate.

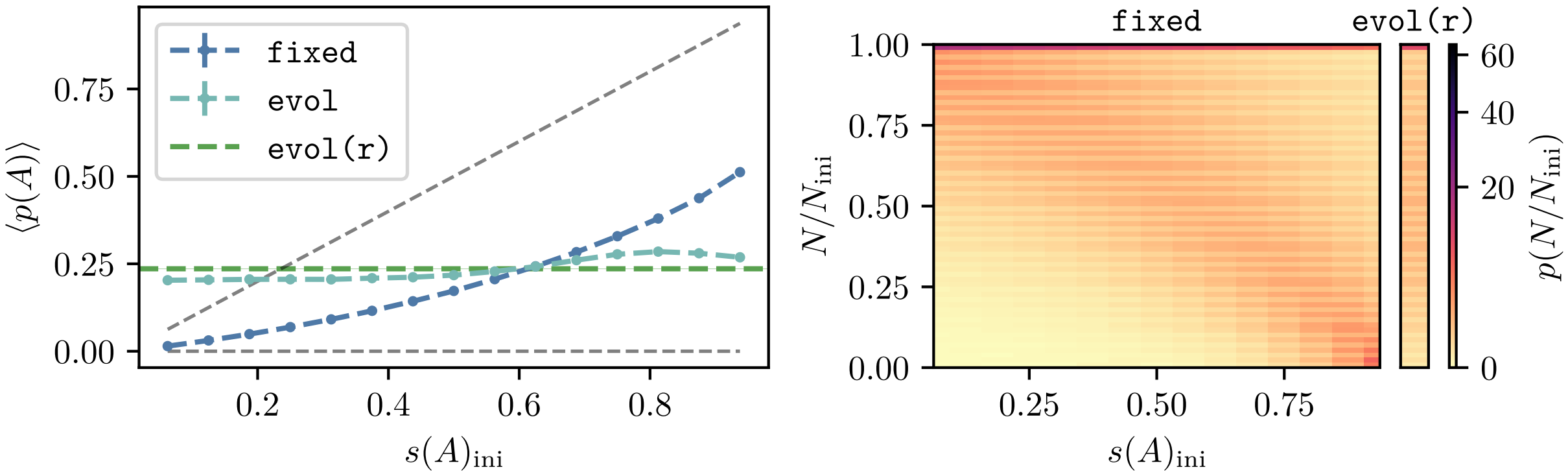

FIG. S2. Average phenotype distribution $\langle p(A) \rangle$ (left) and distribution of the relative population size $N/N_{\text{ini}}$ (right) as a function of the initial strategy $s(A)_{\text{ini}}$ for different kinds of simulations with the PR parameters. The dashed lines indicate the analytical equilibrium values of $\langle p(A) \rangle$ in each of the environments (the upper line corresponds to the favorable environment, and the lower line to the unfavorable one).

## III. EVOLUTIONARY AVERAGE EQUATION

Furthermore, Fig. S3 illustrates how the mean growth rate $\langle \mu \rangle$ (left) and the extinction rate $r_{\text{ext}}$ (right) obtained from evolutionary simulations satisfy the relation proposed in Eq. (2) (represented as dashed lines) for various mutation probabilities $p_{\text{mut}}$. Quantitative agreement with the simulation data confirms that, as argued in Sec. V, the long-term evolutionary dynamics can be accurately approximated as a sequential concatenation of fixed strategies.

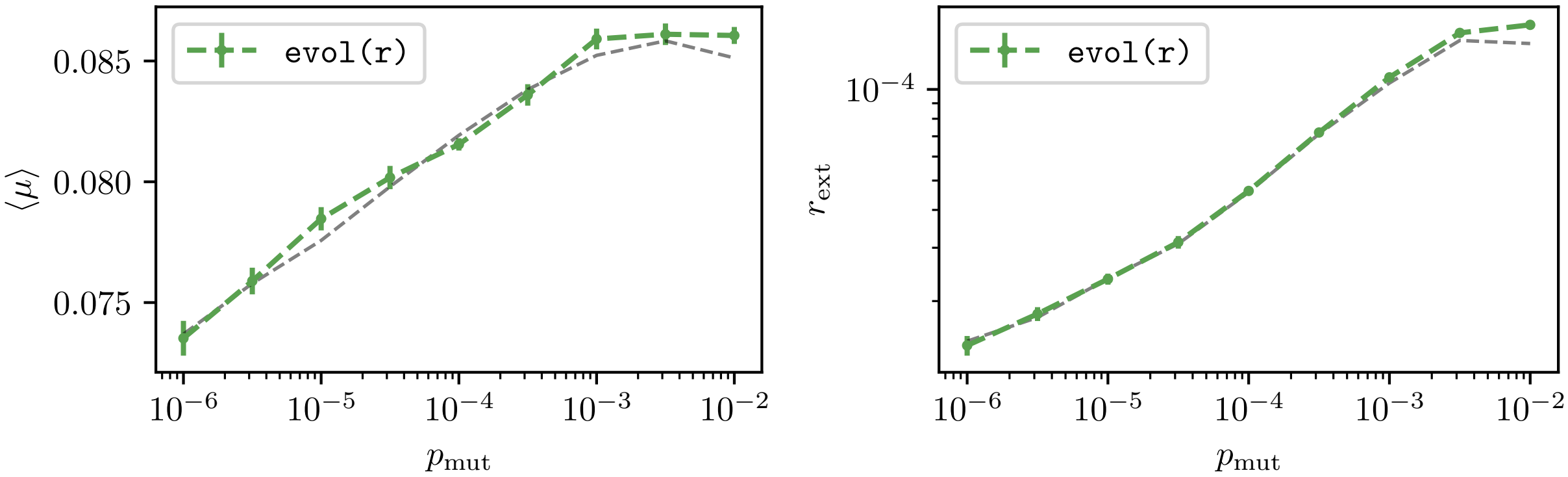

FIG. S3. Mean population growth rate $\langle \mu \rangle$ (left) and extinction rate $r_{\text{ext}}$ (right) as a function of the mutation probability $p_{\text{mut}}$ for the PR parameters in `evol(r)` simulations. The dashed lines indicate the theoretical prediction from Eq. (2).

## IV. HEURISTIC STRATEGY DISTRIBUTIONS

Finally, Fig. S4 and Fig. S5 display the expected (heuristic) and observed (simulated) full distributions of the population-averaged strategy $\bar{s}(A)$ across various initial population sizes $N_{\text{ini}}$ for the PR and SA parameter sets, respectively. For PR parameters (Fig. S4), the agreement between the heuristic prediction and the simulations is remarkably robust, particularly at large values of $N_{\text{ini}}$, whereas the approximation, as expected, degrades in the low-$N_{\text{ini}}$ regime where stochastic fluctuations dominate. In contrast, for the SA parameter set (Fig. S5), a more pronounced quantitative discrepancy emerges; nevertheless, the heuristic approach successfully captures the correct qualitative dependence on $N_{\text{ini}}$, most notably regarding the shifting position of the distribution mean. To further examine the heuristic approximation, Fig. S6 presents the average strategy for the SA parameter set over a wide range of $N_{\rm ini}$ and $p_{\rm mut}$ values. As discussed in Sec. VII, while the heuristic framework reproduces the qualitative trend of the evolutionary outcome, in this case, the system requires significantly higher values of $N_{\rm ini}$ than theoretically predicted to effectively approach the strategy that maximizes the population growth rate $\langle \mu \rangle$.

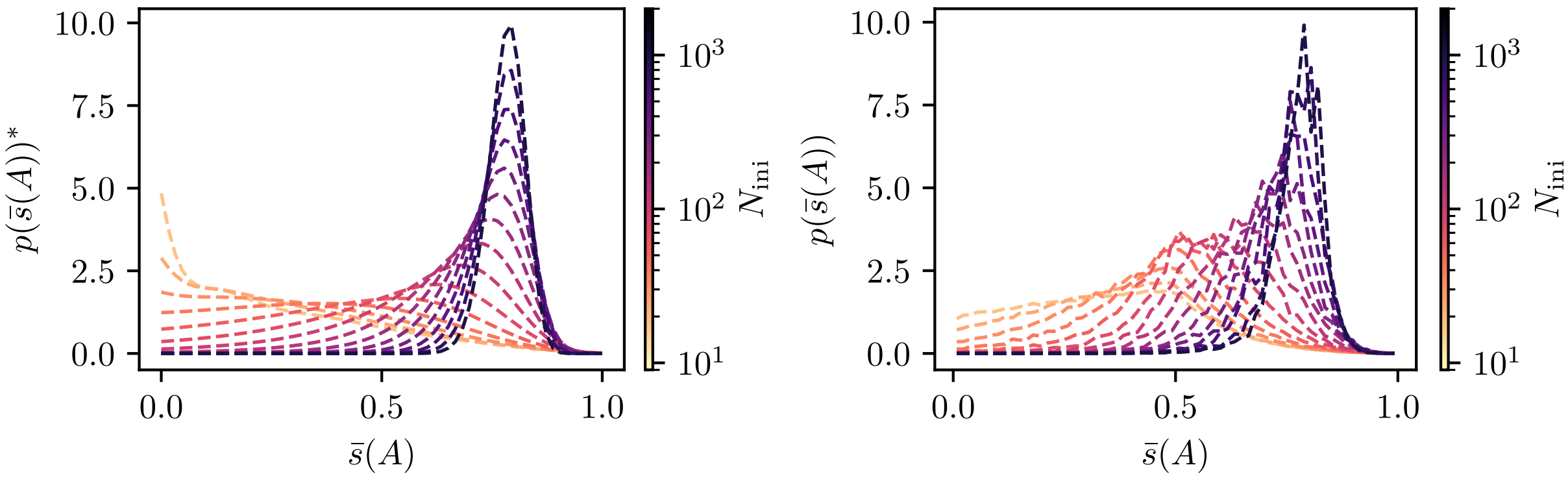

FIG. S4. Distributions of the population-averaged strategy $\bar{s}(A)$ predicted by the heuristic approximation (left) and obtained from `evol(r)` simulations (right) with the PR parameters for different population sizes $N_{\rm ini}$.

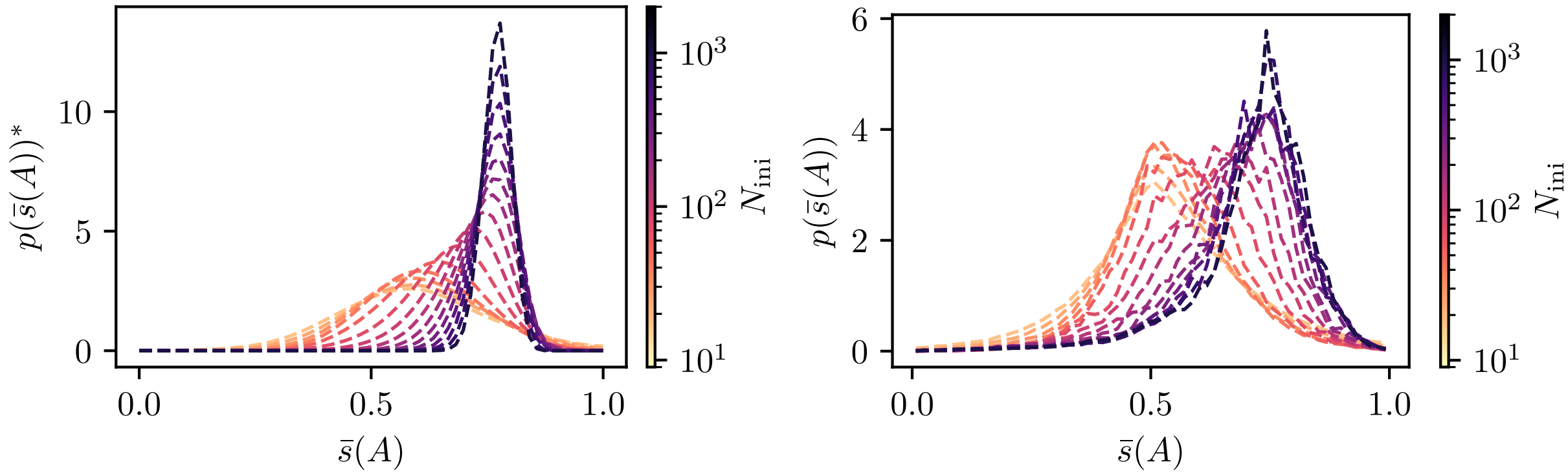

FIG. S5. Distributions of the population-averaged strategy $\bar{s}(A)$ predicted by the heuristic approximation (left) and obtained from `evol(r)` simulations (right) with the SA parameters for different population sizes $N_{\rm ini}$.

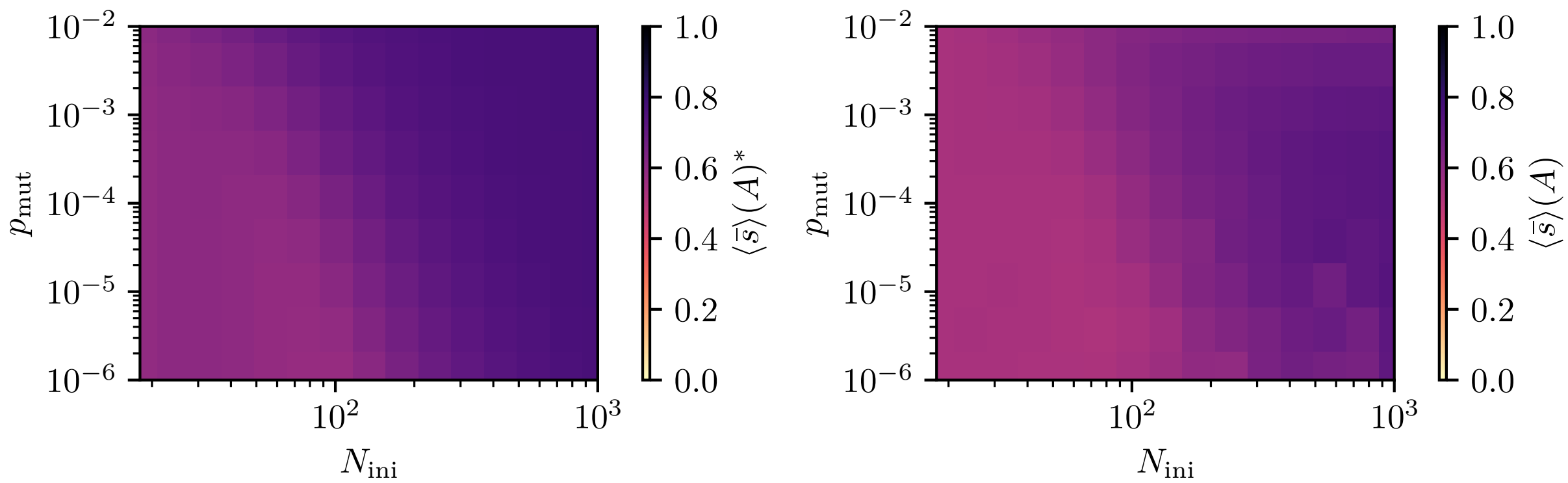

FIG. S6. Expected average strategy from the heuristic distribution $\langle \bar{s}(A) \rangle^*$ (left) and observed average strategy from `evol(r)` simulations (right) for different values of $N_{\rm ini}$ and $p_{\rm mut}$ with the SA parameters.

## V. TIME-DEPENDENT AVERAGE STRATEGY

To complement the time-averaged observables discussed in the main text, we examine the time-dependent population-averaged strategy $\langle\bar{s}\rangle_\tau$. This observable captures the expected average strategy across surviving realizations at a specific elapsed time $\tau$, defined as:

$$\langle\bar{s}\rangle_\tau = \frac{\int_r \int_0^{T_r} \bar{s}_r(t)\delta(\tau - t)dt\rho(r)dr}{\int_r \theta(T_r - \tau)\rho(r)dr}, \tag{S2}$$

where $r$ indexes all possible realizations of the population evolution, $\rho(r)$ is the probability density of realization $r$, $T_r$ is its corresponding extinction time, $\bar{s}_r(t)$ is the population-averaged strategy at time $t$ for that realization, and $\theta$ is the Heaviside step function ensuring that only active (non-extinct) trajectories at time $\tau$ contribute to the average. In contrast, the strategy average presented in the main text corresponds to an overall average over both time and realizations:

$$\langle\bar{s}\rangle = \frac{\int_r \int_0^{T_r} \bar{s}_r(t)dt\rho(r)dr}{\int_r T_r\rho(r)dr}. \tag{S3}$$

The two observables are linked by weighting the time-dependent strategy by the survival probability $P(T > \tau)$:

$$\langle\bar{s}\rangle = \frac{\int_0^\infty \langle\bar{s}\rangle_\tau P(T > \tau)d\tau}{\langle T\rangle}, \tag{S4}$$

where $\langle T\rangle$ is the mean extinction time. Evaluating $\langle\bar{s}\rangle_\tau$ serves two key purposes: first, it exposes the explicit time dependence of the strategy; second, it resolves whether the $N_{\text{ini}}$-dependence observed in $\langle\bar{s}\rangle$ is merely an artifact of concatenating trajectories (which implicitly weighs longer-lived realizations more heavily) or a property visible at fixed elapsed times when analyzing individual trajectories from their onset.

Figure S7 summarizes these findings for the PR parameters. In the left panel, we see how `evol` simulations starting from different initial strategies converge to a common point, confirming the robustness of the steady-state strategy mentioned in the main text. The right panel shows $\langle\bar{s}\rangle_\tau$ for `evol(r)` simulations for different values of $N_{\text{ini}}$. Crucially, the dependence on $N_{\text{ini}}$ persists at all time scales, demonstrating that the population limit directly modulates the time evolution of $s$. Therefore, in practical or experimental settings where environmental fluctuations and population constraints persist only over a finite duration, the resulting strategy observed at the end of that time-frame would exhibit a strong $N_{\text{ini}}$-dependence, closely mirroring our main text results. Finally, it is notable that for large $N_{\text{ini}}$, the curves collapse and approach a constant value with $\tau$. For lower $N_{\text{ini}}$, the reached value decreases; at very small $N_{\text{ini}}$, the dynamic is non-monotonic, showing a peak before shifting toward safer (lower) strategies at larger values of $\tau$.

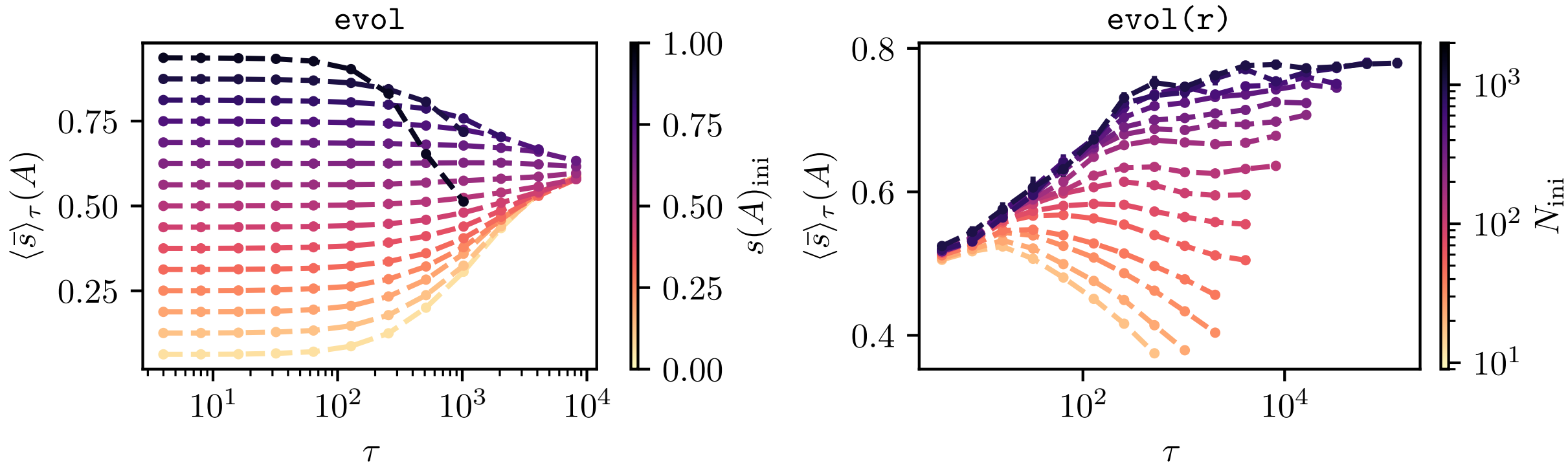

FIG. S7. Time-dependent average strategy $\langle\bar{s}\rangle_\tau$ for the PR parameters. Left: Convergence of `evol` simulations starting from different initial strategies. Right: Dependence of $\langle\bar{s}\rangle_\tau$ on the population limitation for `evol(r)` simulations.